\documentclass[12pt]{iopart}
\usepackage{iopams}  
\usepackage{orcidlink}
\usepackage{amsmath}
\usepackage{amssymb}
\usepackage{latexsym}
\usepackage{bm}
\usepackage{ulem}
\usepackage{graphicx}
\usepackage{ulem}
\usepackage{color}
\usepackage{framed}
\usepackage{dsfont}
\usepackage{amscd}
\usepackage{hyperref}
\usepackage{blindtext}
\usepackage[toc,page]{appendix}
\usepackage{soul}
\usepackage{cancel}
\hypersetup{
    colorlinks=true,
    linkcolor=blue,
    filecolor=magenta,      
    urlcolor=cyan,}
\usepackage{slashed}
\usepackage{cite}

\newcommand{\la}{\langle}
\newcommand{\ra}{\rangle}
\newcommand{\be}{\begin{equation}}
\newcommand{\ee}{\end{equation}}
\newcommand{\p}{\partial}

\newcommand{\half}{\frac{1}{2}}

\newcommand{\ham}{\widehat{\mathsf{H}}}

\newcommand{\vc}{\widehat{\mathsf{c}}}
\newcommand{\vn}{\widehat{\mathsf{n}}}
\newcommand{\vR}{\mathbf{R}}

\begin{document}

\title{Chiral symmetry breaking  and topological charge of  zigzag graphene nanoribbons}


\author{
	Hyun Cheol Lee$^{1}$\orcidlink{0000-0003-1352-8851}, 
	and S.-R. Eric Yang$^{2}$\orcidlink{0000-0003-3377-1859}}
\address{$^1$Department of Physics, Sogang University 04107, Seoul, Korea}
\address{$^2$Department of Physics, Korea  University, Seoul 02841, Korea}
\ead{hyunlee@sogang.ac.kr, eyang812@gmail.com}

\begin{abstract}
 { Interacting quasi-one-dimensional  zigzag graphene nanoribbons display gapped edge excitations.}
Although the self-consistent Hartree-Fock fields break chiral symmetry, our work demonstrates that zigzag graphene nanoribbons maintain  their status as short-range entangled symmetry-protected topological insulators. The relevant symmetry involves combined mirror and time-reversal operations.  In undoped ribbons displaying edge ferromagnetism, the band gap edge states with a topological charge form  on the  zigzag edges.   An analysis of the anomalous continuity equation elucidates that this topological charge is induced by the gap term.
In low-doped zigzag ribbons, where the ground state exhibits edge spin density waves, this topological charge appears   as a nearly zero-energy edge mode. Our system is outside the conventional calssification for topological insulators.

\end{abstract}

\section{Introduction}
\label{sec:intro}


  Non-interacting zigzag  graphene  nanoribbons\cite{Brey,Yang} exhibit chiral symmetry and are topological insulators\cite{Pachos2009}. However, this symmetry breaks down  in the presence of electron interactions, leading to uncertainty regarding the definition of a topological charge.
 	Our paper aims to elucidate the formation of a topological charge in disorder-free interacting zigzag graphene nanoribbons.
 We will demonstrate that in the presence of such interactions, the relevant symmetry combines mirror and time-reversal operations. The breakdown of chiral symmetry leads us to utilize axial current instead of ordinary current for an accurate definition of a topological charge, mirroring the approach by Goldstone and Wilczek \cite{Goldstone} in the context of solitons in polyacetylene \cite{Heeger}.
 This issue holds experimental significance due to the recent fabrication of atomically precise nanoribbons (Cai et al. \cite{Cai2}, Kolmer et al. \cite{Kolmer}, Houtsma et al. \cite{Houtsma2021}).

One way to define  a topological charge  is through the generation of mass (or gap), which breaks chiral symmetry\cite{Goldstone}.
	 This concept can be elucidated using a one-dimensional Dirac equation.
In one spatial dimension (and generally for all odd spatial dimensions), the gapless (or massless in the particle physics context) Dirac fermions possess the property of \textit{chirality}. The Lagrangian density of the massless Dirac fermion  (see the Appendix for details and conventions) is invariant under \textit{two} kinds of global gauge transformations:
\be
\psi \to e^{i \Lambda} \psi \;\; 
\text{ordinary}, \quad \psi \to e^{i \gamma^5 \Lambda} \psi \;\;\text{axial or chiral},
\ee
where $\psi$ is a Dirac spinor, and  $\gamma^5 = \gamma^0 \gamma^1$. Chirality is defined to be the eigenvalue of $\gamma^5$, which can take only $\pm 1$ values.
According to classical Noether's theorem, these invariances imply the conservation of the corresponding currents (Fermi velocity may be multiplied to the spatial component):
\be
\label{1dcurrent}
j^{\mu}_{\mathrm{ordinary}} =  \bar{\psi} \gamma^\mu  \psi, \quad 
j^{\mu 5}_{\mathrm{axial}} = \bar{\psi} \gamma^\mu \gamma^5 \psi.
\ee
Owing to a special property of gamma matrices in two dimensions, the above two currents can be related by:
\be
j^{\mu 5} = - \epsilon^{\mu \nu} j_\nu,
\ee
where $\epsilon^{\mu \nu}$ is the two-dimensional Levi-Civita symbol with $\epsilon^{01}=1$.
Expressing the Dirac spinor $\psi$ in $\gamma^5$-diagonal basis as 
$(\psi_R,\psi_L)^{\rm t}$ ('t' is transpose),
the explicit expressions of currents  are given by (note that $j_1 = -j^1$, etc.):
\be
j^0 = \psi^\dag_R \psi_R + \psi^\dag_L \psi_L =j^{1,5} ,
\quad j^1 = \psi^\dag_R \psi_R - \psi^\dag_L \psi_L= j^{0,5}.
\ee

A mass term for Dirac fermion couples $\psi_R$ and $\psi_L$. 
Two independent mass terms (in Hamiltonian) are allowed in the following way:
\be
\label{masses}
m (\psi^\dag_R \psi_L + \text{h.c.} ) + i m ' ( \psi^\dag_R \psi_L - \text{h.c.} ).
\ee
The above mass terms break the chiral symmetry explicitly.
 From the viewpoint of a one-dimensional polyacetylene system, these masses correspond to
 the even and the odd components  of dimerization. In the presence of the mass terms, the continuity equation for the axial current is modified as follows \cite{Goldstone,Fradkin}:
\be
\label{anomaly_massive}
\p_\mu j^{\mu 5} =
\frac{\p j^1}{\p t}+ \frac{\p j^0}{\p x}=  - 2 (m- i m') \psi_R^\dag \psi_L - 2 (m + i m') \psi_L^\dag \psi_R
\equiv \mathcal{M}.
\ee
Eq.(\ref{anomaly_massive}) remains valid even if the mass parameters depend on coordinates. For slow adiabatic temporal changes, the time derivative of Eq.(\ref{anomaly_massive}) can be neglected, and we have ($\rho=j^0$ denotes charge density).
\be
\label{anomaly_massive2}
 \frac{\p\rho }{\p x}= \mathcal{M}(x).
\ee

An abrupt step-like change of the mass terms can give rise to a solitonic\cite{Rebbi, Heeger} topological charge\cite{Goldstone} accumulated at the soliton location, where 
the spectral weight of a soliton is equally divided between  conduction and valence bands.
More precisely, $\mathcal{M}(x)$ in Eq.(\ref{anomaly_massive}) corresponds to the topological charge\cite{GV2019,Pach} density after subtracting out the trivial vacuum contribution\cite{Goldstone}.  Then, the integration of Eq.(\ref{anomaly_massive2}) over some interval with endpoints (say, $[-\infty,\infty]$) gives the topological charge $Q_{\rm top}$ associated with the solitonic configuration of mass parameters
\be
\label{topo-accumulation}
\rho(\infty)-\rho(-\infty)=\int_{-\infty}^{\infty} dx \; \mathcal{M}(x) = Q_{\rm top}.
\ee
The determination of this topological charge relies on the topological aspects of the mass parameters, defined as $\mathcal{M}(x)=\frac{\partial \theta}{\partial x}$, where the order parameter $\theta(x)$ represents a kink:
\be
 \theta(x)=\begin{cases}
	-1/2 & \text{for }x<0\\    
	+1/2  &   \text{for }x>0.
\end{cases}
\ee
 In polyacetylene $ \theta(x)$ represents dimerization order\cite{Heeger}.
However, this topological charge  is not necessarily quantized in general circumstances\cite{Goldstone}. Then, by Eq.(\ref{topo-accumulation}), the induced fermion number can be identified with the topological charge $Q_{\rm top}$ of mass parameters, which is in turn determined by their boundary behavior\cite{Rebbi}.

\begin {figure}[!hbpt]
\begin{center}
	\includegraphics[width=0.4\textwidth]{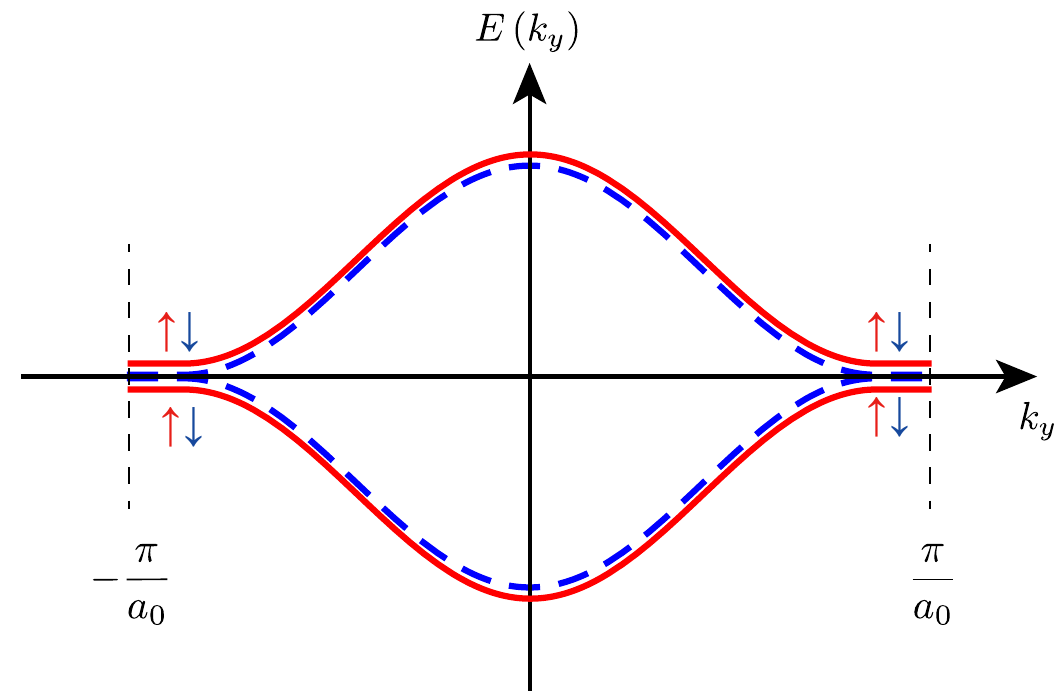}
	\includegraphics[width=0.4\textwidth]{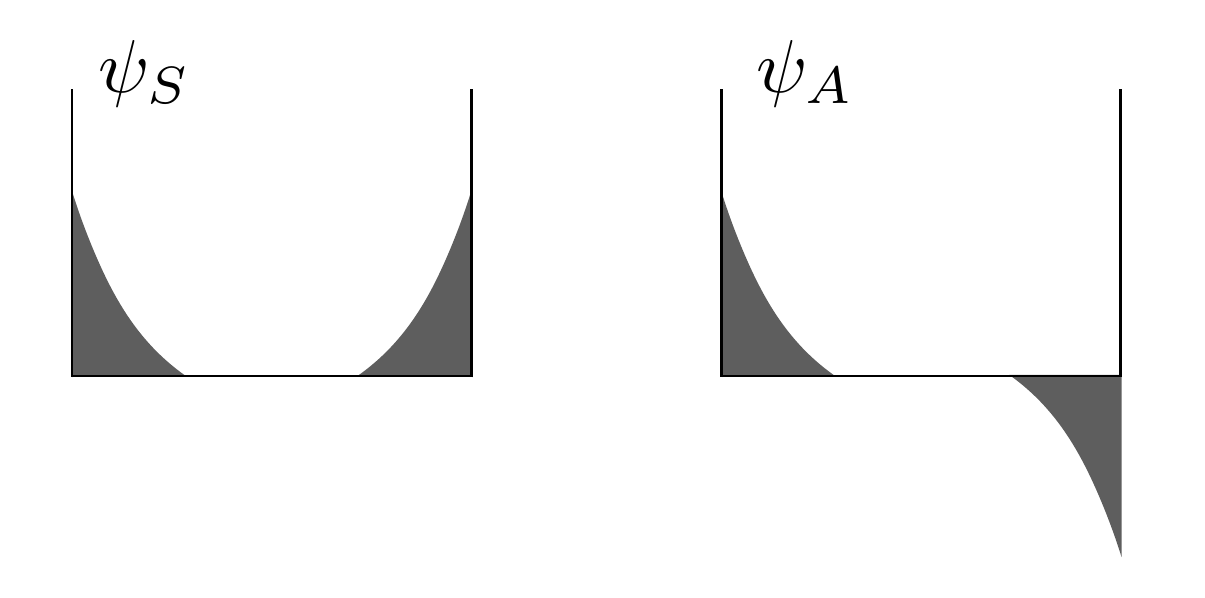}
	\caption{\textbf{First}: The band structure of a non-interacting zigzag nanoribbon displays nearly zero-energy states localized on the zigzag edges.
		\textbf{Second} : Solitonic nearly zero-energy states appear near $k=\pm \pi/a_0$, comprising symmetric $\psi_B$ and antisymmetric $\psi_A$ states well-localized on the left and right zigzag edges ($a_0$ represents the unit cell length of the zigzag ribbon). }
	\label{band2}
\end{center}
\end{figure}

\begin {figure}[!hbpt]
\begin{center}
	\includegraphics[width=0.3\textwidth]{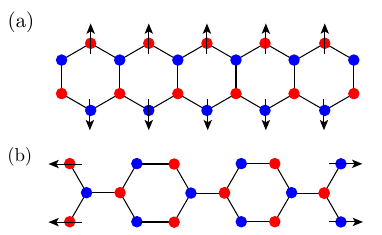}
	\includegraphics[width=0.7\textwidth]{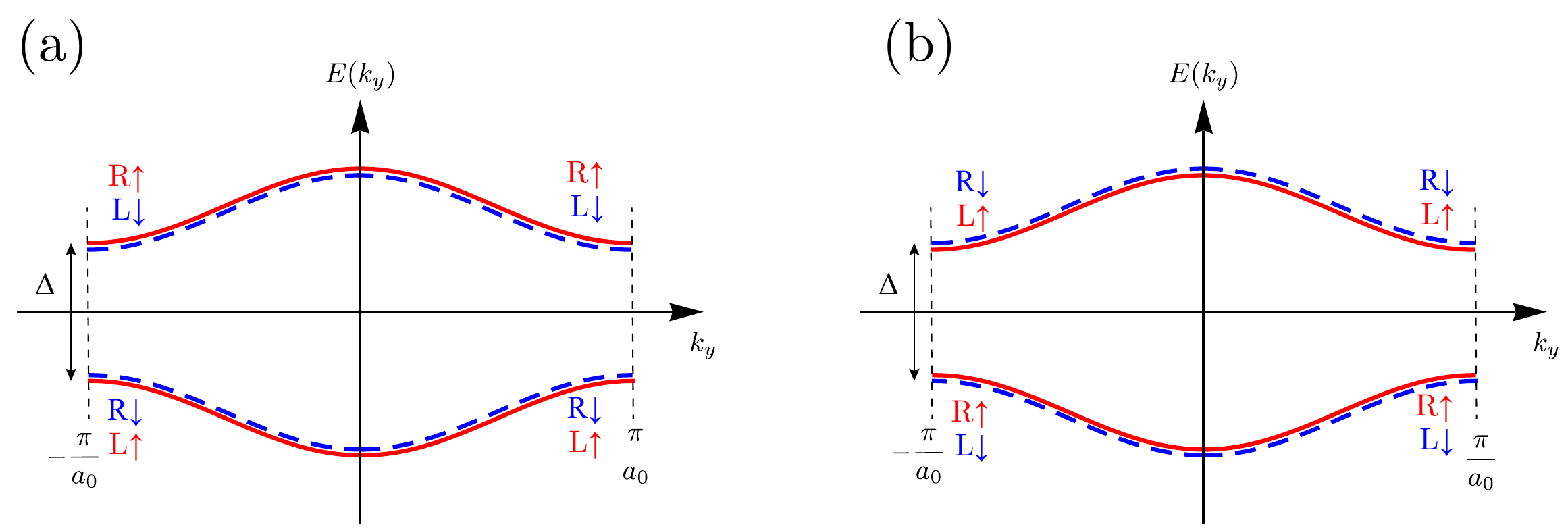}	
	\includegraphics[width=0.4\textwidth]{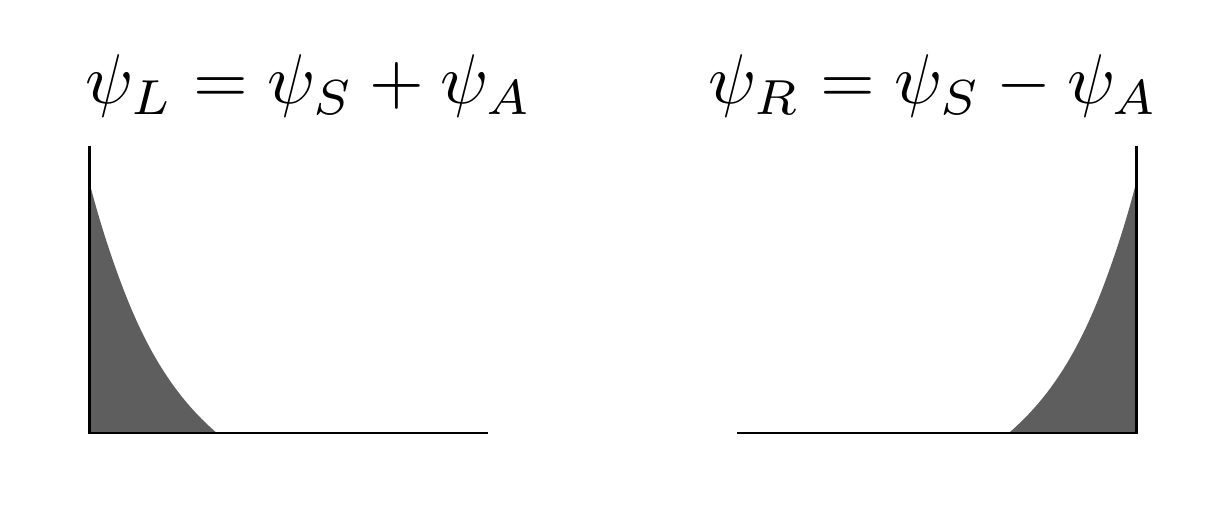}
	\caption{\textbf{First}: Two interacting  rectangular nanoribbons are shown. A graphene  zigzag nanoribbon (a) and an  armchair nanoribbon (b) consist of two sublattices, A (red sites) and B (blue sites).   In an zigzag (armchair)  ribbon, the left and right carbon sites are part of the armchair (zigzag) edges. The net site spins of the zigzag edges are displayed. The spins are 'down' on the red 'A' sites and 'up' on the blue 'B' sites. Zigzag edges are antiferromagnetically coupled, while each of them is ferromagnetically ordered. Site spin values are determined by the occupation numbers $n_{i\sigma}$: $s_{iz}=\frac{1}{2}[n_{i\uparrow}-n_{i\downarrow}]$, where $i$ denotes a site,  and $\sigma$ represents spin up or down.  A edge sites are predominantly occupied by spin-down, while the B edge sites are predominantly occupied by the opposite spin. Therefore, in each sublattice, the total occupation numbers of spin-up and spin-down electrons are different.
	\textbf{Second}: Band structures of two degenerate ground states of a zigzag nanoribbon in the presence of on-site Coulomb repulsion. Zigzag edge states are found near the band gap edges, occurring at energies approximately $E\approx \pm\Delta/2$, with momenta around $k\approx \pm\pi/a_0$.  L and R stand for left and right edge localized states.
	\textbf{Third}: Zigzag edge states $\psi_L$ and $\psi_R$, exclusively localized on either the left or right edge, result from the combination of symmetric $\psi_S$ and antisymmetric $\psi_A$ solitonic states, as depicted in Fig.\ref{band2}.	}
	\label{band1}
\end{center}
\end{figure}

Our goal in this paper is to try to understand the formation of a topological charge in disorder-free {\it interacting} zigzag graphene nanoribbons in view of the above statement. 
First, let us give a brief introduction to non-interacting graphene nanoribbons.  The band structure of a non-interacting zigzag graphene nanoribbon  is shown in Fig.\ref{band2}.  Zigzag graphene  nanoribbons\cite{Neto,Brey2006,Yang,yang1} consist of A and B sublattices, as seen in the top figure of Fig.\ref{band1}.  The A/B sublattice degree of freedom is referred to as the chirality. The chiral transformation\cite{Ryu2002,Delplace2011} on  electron operators is implemented by
\be
\label{chiral-transform}
\widehat{\mathsf{a}}_i\rightarrow \widehat{\mathsf{a}}_i,  \quad   \widehat{\mathsf{b}}_i\rightarrow -\widehat{\mathsf{b}}_i,
\ee
where $\widehat{\mathsf{a}}_i$ and $\widehat{\mathsf{b}}_i$ are the electron destruction operators of A and B sublattice sites which are similar to the Dirac fermion
$\psi_R, \psi_L$ under the action of $\gamma^5$ (see the Appendix).
Under the transformation Eq.(\ref{chiral-transform}), the nearest-neighbor hopping term between A-B sublattices of the tight-binding Hamiltonian of graphene nanoribbon without on-site interaction changes sign, and this is called the chiral symmetry. 
This chiral symmetry implies a particle-hole symmetry within the energy spectrum and ensures the existence of zero-energy soliton\cite{Sasaki2010, ChiralTH}  edge states.
Thus, non-interacting graphene zigzag nanoribbons are deemed a symmetry-protected topological insulator\cite{Hasan2010,Wen11}. 

Now, the inclusion of on-site repulsion interactions among electrons explicitly breaks chiral symmetry, as the density operator does not alter its sign under the transformation described by Eq.(\ref{chiral-transform}). This interaction introduces an energy gap, $\Delta$.  Despite this energy gap, the persistence of edge states is evident, as depicted in Fig.\ref{band1}. Furthermore, interacting zigzag graphene nanoribbons manifest as short-range entangled  symmetry-protected topological insulators\cite{Wen11}. 
	 (Here, we exclude consideration of disordered zigzag ribbons that exhibit long-range entangled\cite{Yang2019,Yang2021} topological order\cite{Wen11}.)
Consequently, it is natural to inquire about the topological origin of these edge states in the presence of on-site interactions. (In this paper, we do not consider armchair ribbons since the physical origin of a gap may differ.)


We aim to address the aforementioned questions by investigating the interacting nanoribbons within the framework of the Hartree-Fock (HF) approach and employing an anomalous continuity equation similar to Eq. (\ref{anomaly_massive}).
The self-consistent HF fields break chiral symmetry. However, the self-consistent fields of the opposite zigzag edges exhibit antisymmetry under mirror symmetry operation. The relevant symmetry combines mirror and time-reversal operations, enabling the formation of a topological charge induced by the mass on ferromagnetic zigzag edges in undoped ribbons. This mechanism is analogous to the equation in Eq.(\ref{topo-accumulation}) and is of a non-perturbative nature.
This topological charge persists even in the low-doped region where the ground state displays edge spin density waves.   It  appears as a nearly zero energy edge mode.  Our work demonstrates that interacting graphene nanoribbons retain their status as symmetry-protected topological insulators.  {  We believe that the conventional classification scheme of topological insulators ~\cite{Andreas2008,Ryu_2010}  does not directly apply to zigzag graphene nanoribbons because interacting zigzag ribbons display gapful edge excitations, whereas the conventional classification applies to gapless excitations.   Moreover, the quasi-one-dimensionality of zigzag ribbons is important as it spatially separates the opposite zigzag edges of different chiralities.}


\section{Tight-binding Hubbard model of  Nanoribbon}
\label{sec:tightbinding}

 Let us consider the narrowest armchair nanoribbon with zigzag edges and the narrowest zigzag nanoribbon, as illustrated in the top figure of Fig. \ref{band1}.
The Hamiltonian of the  tight-binding Hubbard model of these nanoribbons\cite{Fujita, Pisa1,Stau,Yang2022} consists of the hopping term
with hopping amplitude $t$ and the Hubbard term for 
on-site Coulomb repulsion $U$
\be
\label{hamil-tb}
\ham=\ham_{t} + \ham_{U}.
\ee
The second quantized hopping Hamiltonian is (h.c. is the Hermitian conjugate, 
and  $\sigma = \uparrow,\downarrow$ is spin)
\be
\label{hoppingH}
\ham_{t} = - t \sum_{< \vR, \vR' > \sigma}  ( \vc^\dag_{\vR \sigma } \vc_{\vR' \sigma} + \mathrm{h.c.} ),
\ee
where $\vc_{\vR,\sigma}$ is the electron destruction operator at site $\vR$
with spin $\sigma$ and 
$< \vR, \vR' > $ denotes a nearest-neighbor pair. One of the pair belongs to A-sublattice, while 
the other is in B-sublattice.
The Hubbard term is given by
\be
\label{hamil:hubbard}
\ham_{U} = U \sum_{\vR } \vn_{\vR \uparrow} \vn_{\vR \downarrow},
\ee
where $\vn_{\vR \sigma}=\vc^\dag_{\vR \sigma} \vc_{\vR \sigma}$ is the electron number operator at site $\vR$ 
with spin $\sigma$.
As discussed in Section \ref{sec:intro}, $\ham_t$ posssses the chiral symmetry, while 
the 
$\ham_{U}$ explicitly breaks it.

Applying HF mean-field theory to Eq.(\ref{hamil:hubbard}) at \textit{half-filling}, 
a quadratic mean-field Hamiltonian $\ham_{\text{HF}}$ can be obtained:
\be
\label{HF_hamil}
 \ham_{\text{HF}} = \ham_{t} + \ham_{U,{\rm HF}}, \quad 
\ham_{U,{\rm HF}} = \sum_{\vR \sigma}   \,  \Delta_{\vR \sigma}  \vn_{\vR \sigma},
\ee
where the gap parameter $\Delta_{\vR \sigma}$, which corresponds to the 
mass parameter of Eq.(\ref{anomaly_massive}), is defined by
\be
\label{gap-parameter}
\Delta_{\vR \sigma} = U \big(  \la \vn_{\vR \bar{\sigma}} \ra -\frac{1}{2} \big ),\quad  
\bar{\sigma} = \text{the opposite spin of } \; \sigma.
\ee
A constant term has been added to $\ham_{\text{HF}} $ to make the Fermi energy to be zero.
 The site gap $\Delta_{\vR \overline{\sigma}}$ associated with spin $\overline{\sigma}$ can be understood as the staggered potential energy for an electron with the opposite spin $\sigma$. This potential is staggered since it changes sign between A- and B-sites, as depicted in Fig. \ref{mirror-x}.
It is important to note that the mean-field HF Hamiltonian can be split into
the spin-up and the spin-down components \cite{Yang2020}:
\be
\label{hamil}
\ham_{\text{HF}}=\ham_{{\text{HF}} \uparrow} + \ham_{{\text{HF}} \downarrow}.
\ee

The band structure \textit{without} the Hubbard interaction $\ham_U$ taken into 
account  was shown in Fig.\ref{band2}.  
The nearly zero-energy edge states exist close to the Brillouin zone boundary, and they exhibit non-magnetic behavior (the net spin polarization of a zigzag edge is zero).
The existence of these \textit{zero} energy chiral edge states crucially 
rests on the chiral symmetry.
However, a self-consistent solution of the HF mean-field Hamiltonian reveals that for ribbon widths shorter than $\text{100 \AA}$, the ground state is antiferromagnetic { in the undoped case.  Local density approximation calculations \cite{Pisa1,ShiHuaTan2014,Lyang} also yield similar results.}  In this state, site spins of sublattice A (B) are spin-up (-down), and the state is doubly degenerate with an energy gap.
The chiral edge states \textit{do} exist, and 
these chiral edge states are {now oppositely polarized (edge antiferromagnetism) \cite{Fujita}. 
Each zigzag edge displays ferromagnetism but the opposite zigzag edges with different chirality are antiferromagnetically coupled (see  the top figure of Fig. \ref{band1}).
The band structures are displayed in the second figure in Fig. \ref{band1}.  { The HF results qualitatively agree with those of density matrix renormalization methods ~\cite{Yang2022}, both in doped and undoped zigzag ribbons. In the low-doped regime, it is found that edge spin density waves develop in contrast to the ferromagnetic edges observed in the undoped case.  Thus, doping is a singular effect.
 A staggered potential may be applied  to graphene nanoribbons~\cite{Jeong2017}.  In this case, mirror and time reversal symmetries are broken, as is their combined operation. The HF method shows that the charge of each zigzag edge may assume non-integer values instead of integer values.
}


\section{Symmetry of  undoped rectangular ribbons}
\label{sec:hidden}

Ribbons exhibit a symmetry based on the combined operation of parity transformation (mirror reflection) and time reversal transformations, as described by the relationship:
\be
\label{pattern}
\la \vn_{\vR \sigma} \ra = \la \vn_{\overline{\vR} \bar{\sigma}} \ra.
\ee
This relation is illustrated in Fig.\ref{PH}.      Note that $\overline{\vR}$ is the mirror symmetric site of $\vR$ and $\bar{\sigma}$ is the opposite spin of $\sigma$.  Here, the relevant  mirror axis of the zigzag nanoribbon is the \textit{ horizontal} x-axis and 
that of the armchair nanoribbon the \textit{ vertical} y-axis through the center of the ribbon.
This symmetry aligns with the existence of a spin-up edge state on one zigzag edge and the corresponding spin-down edge state on the opposite zigzag edge.

Then the definition of the gap parameter Eq.(\ref{gap-parameter}) immediately implies that 
\be
\label{pattern2}
\Delta_{\vR \sigma} = U \big ( \la \vn_{\vR \bar{\sigma}} \ra-\half \big ) =
U \big ( \la \vn_{\overline{\vR} \sigma} \ra-\half \big ) = \Delta_{\overline{\vR} \bar{\sigma}}.
\ee
Note that Eq.(\ref{pattern2}) is valid even for the low-doped case as long as Eq.(\ref{pattern}) is valid.
It is important to note that this mirror symmetry is accompanied by spin reversal and that
the symmetry in the occupation numbers involves a spin reversal but not in the site gap values.  At half-filling (undoped case),  the condition $\la \vn_{\vR \sigma} \ra + \la \vn_{\vR \bar{\sigma}} \ra = 1$ being combined with Eq.(\ref{pattern})  yields 
\be
\label{pattern3}
\Delta_{\vR \sigma} = - \Delta_{\overline{\vR} \sigma},
\ee
which is demonstrated in Fig. \ref{mirror-x}.

\begin {figure}[!hbpt]
\begin{center}
	\includegraphics[width=0.5\textwidth]{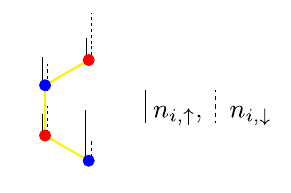}
	\caption{ The values of $n_{i,\uparrow}$ (represented by vertical solid lines) and $n_{i,\downarrow}$ (shown as vertical dashed lines) within a unit cell of the shortest-width zigzag ribbon are depicted.  Note that for each site  $ \la n_{\vR,\uparrow} \ra+ \la n_{\vR,\downarrow} \ra=1$. 
	}
	\label{PH}
\end{center}
\end{figure}

\begin {figure}[!hbpt]
\begin{center}
	\includegraphics[width=0.7\textwidth]{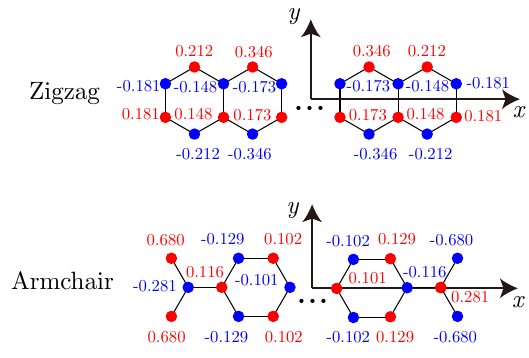}
	\caption{Magnitude of the self-consistent  gap $\Delta_{\vR \uparrow} $ at each site. Mirror symmetry about the horizontal (vertical) axis is present only in the armchair (zigzag) case.  We will call these axes the mirror axes. Note that \textit{only up-spin} gap parameters are indicated.   }
	\label{mirror-x}
\end{center}
\end{figure}

\section{Symmetry of low-doped zigzag ribbons}

Let us consider zigzag ribbons in the low doping limit.   It turns out that doping has a singular effect\cite{Yang2022}: the grounds state displays an edge spin density wave, as shown in Fig.\ref{DopedZig}.  This is in contrast to the uniform edge  ferromagnetism of the half-filled case.   Note that in the low-doped region, the electron spins are collinear to a very good approximation, as confirmed by the density matrix renormalization group approach in the matrix product representation. This implies that our Hamiltonian at half-filling may still be applicable\cite{Yang2022}.

\begin {figure}[!hbpt]
\begin{center}
	\includegraphics[width=0.7\textwidth]{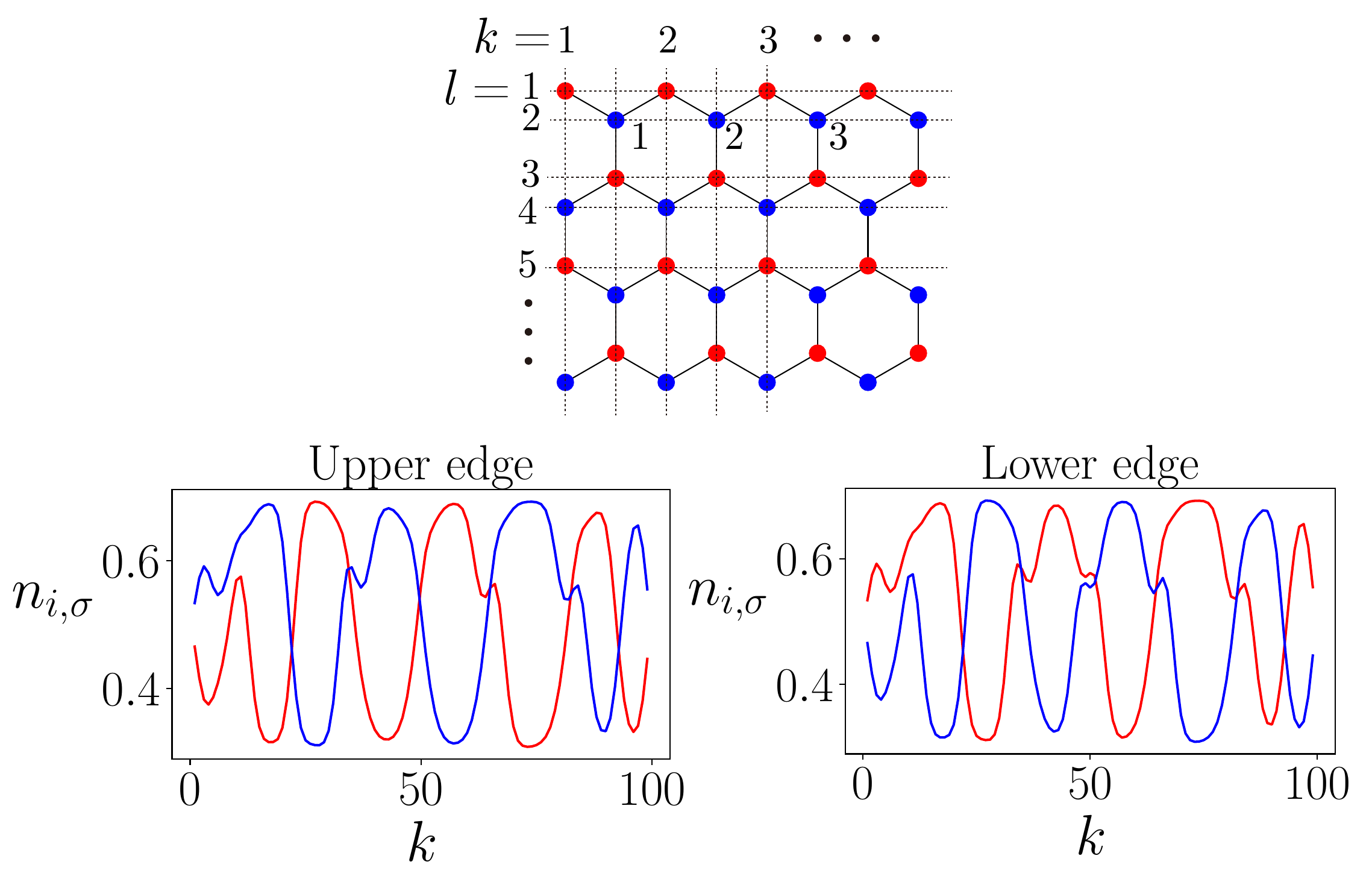}
	\caption{Spin-up and spin-down profiles of $\la n_{\vR,\sigma} \ra $ at the zigzag edges of a doped zigzag ribbon with an additional $\delta N=12$ electrons added to  half-filling.  Note that $\la n_{\vR,\uparrow}\ra + \la n_{\vR,\downarrow} \ra \neq \text{constatnt}$, i.e., the varies from site to site.  The site spins  $s_{ \vR, z}=\frac{1}{2}[\la n_{\vR,\uparrow} \ra
	-\la n_{\vR,\downarrow} \ra ]$ of the upper edge is out of phase with that of the lower edge.  
	The site $\vR$ is parametrized by $(k,l)$.
	}
	\label{DopedZig}
\end{center}
\end{figure}

To understand the role of the doping we write 
the filling condition as 
\be
\label{filling-condition}
\la \vn_{\vR \sigma} \ra + \la \vn_{\vR \bar{\sigma}} \ra = 1+\delta,
\ee
where $\delta$ is the doping level. Then we have
\be
\label{filling-effect}
\Delta_{\vR \sigma}  = U \big ( \la \vn_{\vR \bar{\sigma}} \ra-\half \big )
=U \big (1+\delta - \la \vn_{\vR \sigma} \ra-\half \big )
=- \Delta_{\vR \bar{\sigma}}+ U \delta=- \Delta_{\overline{\vR} \sigma}+U \delta.
\ee
Eq.(\ref{filling-effect}) shows that the antisymmetry of the gap parameter $\Delta_{\vR \sigma}$ observed at half-filling in Fig. \ref{mirror-x} is disrupted when moving away from the half-filling state relative to the mirror axis.


\section{Anomalous continuity equation}
\label{sec:continuity}

The total current operator of the tight-binding Hamiltonian Eq.(\ref{hamil-tb}) is given by
\be
\label{currentoperator}
\mathbf{J}_{\text{tot}}= i  t \sum_{ <\mathbf{R}, \mathbf{R}'>, \sigma} (\mathbf{R}-\mathbf{R}') \,
(\vc^\dag_{\mathbf{R}, \sigma}  \vc_{\mathbf{R}', \sigma} - \text{h.c.} ),
\ee
which is defined such that 
the ordinary continuity equation is satisfied.
Remember, in one-dimensional Dirac fermions, the axial current can be defined, where the roles of charge density $\rho$ and spatial charge current density $j$ interchange between ordinary and axial current (as shown in Eq.(\ref{1dcurrent})).  Additionally, the Fr\"olich equation governing the (unpinned) sliding charge density wave \cite{Frolich} is expressed as:
\be
\label{frolich}
\nabla \rho + \frac{1}{v_F^2} \frac{\p j}{\p t} = \kappa E.
\ee
Here, $\kappa$ represents compressibility, $v_F$ signifies Fermi velocity, and $E$ denotes an electric field. In one-dimensional systems, Eq.(\ref{frolich}) essentially embodies the well-known axial anomaly\cite{Fradkin}. These insights lead us to consider the \textit{time derivative} of the spatial current, as outlined in Eq.(\ref{currentoperator}).
This approach is motivated by the pronounced manifestation of chiral (axial) symmetry breaking effects, notably evident in the continuity equation for the axial current (Eq.(\ref{anomaly_massive})).

The time derivative can be found from the commutator with $\ham_{\text{HF}}$ Eq.(\ref{HF_hamil}):
\be
\label{timed-ano}
\frac{\p \mathbf{J}_{\text{tot}}}{\p t}=
\frac{i}{\hbar}[\ham_{U,\text{HF}},\mathbf{J}_{\text{tot}}]
+\frac{i}{\hbar}[\ham_{t},\mathbf{J}_{\text{tot}}].
\ee
This expression comprises two contributions: one from the HF decomposed Hubbard interaction and the other from the hopping term.
Taking the HF ground state expectation value of Eq.(\ref{timed-ano}) and considering the temporal adiabatic limit, $(-i \hbar) \frac{\partial \langle \mathbf{J}_{\text{tot}} \rangle}{\partial t} \approx 0$ we find :
\be
\label{expectationvalue}
(-i \hbar) \frac{ \p \la \mathbf{J}_{\text{tot}} \ra } {\p t}   =
   \la [\ham_{U,\text{HF}},\mathbf{J}_{\text{tot}}] \ra +  \la [\ham_{t},\mathbf{J}_{\text{tot}}]  \ra\approx 0.
\ee
Then,  we explore the implications of Eq.(\ref{expectationvalue}) in light of the (integrated) anomalous continuity equation (Eq.(\ref{anomaly_massive})) and the accumulation of the topological charge (Eq.(\ref{topo-accumulation})).

The computation of the commutators are straightforward. For each spin projection, we find
\be
\label{gapoperator}
[\ham_{U,\text{HF}},\mathbf{J}_{\text{tot},\sigma}]  =
(+i)  t \sum_{ < \mathbf{R}, \mathbf{R}' > } (\mathbf{R}-\mathbf{R}')
( \Delta_{\vR \sigma}-\Delta_{\vR' \sigma} )  \big (\vc^\dag_{\mathbf{R}, \sigma} \vc_{\vR' \sigma}+
\text{h.c.} \big ),
\ee
and
\be
\label{densitygradient}
[\ham_{t},\mathbf{J}_{\text{tot},\sigma}]
=(-i) t^2\sum_{<\vR_1,\vR>} 
  \sum_{ <\mathbf{R}, \mathbf{R}'>} (\mathbf{R}-\mathbf{R}') \big( \vc^\dag_{\vR_1 \sigma} \vc_{\mathbf{R}' \sigma}+\text{h.c.} \big ).
\ee
It's important to clarify that in Eq.(\ref{densitygradient}), $\vR_1$ and $\vR'$ are \textit{not} a nearest neighbor pair, see Fig.\ref{hoppingRR}. To elaborate further, if we consider a fixed $\vR'$, then $\vR$ must be a nearest neighbor in the other sublattice, thereby determining the bond vector $\vR - \vR'$. Subsequently, $\vR_1$ should be a nearest neighbor of $\vR$, placing it in the same sublattice as $\vR'$.
Now, considering the proximity of $\vR'$ to $\vR$, $\vR_1$ \textit{can} be $\vR'$, resulting in $\vc^\dagger_{\vR'\sigma} \vc_{\vR' \sigma} + \text{h.c.}$, effectively representing two times the number operator at $\vR'$. However, it's crucial to note that $\vR_1$ might not necessarily be the same as $\vR'$; in such cases, $\vR_1$ becomes a next-nearest neighbor site from $\vR'$ while still being in the same sublattice.
Thus, Eq.(\ref{densitygradient}) can be expressed as
\be
\label{densitygradient1}
[\ham_{t},\mathbf{J}_{\text{tot},\sigma}] =(-i) 2 t^2 \sum_{\vR'} \,
 \sum_{ <\mathbf{R}, \mathbf{R}'>} (\mathbf{R}-\mathbf{R}')   \vn_{\vR' \sigma} +
\text{next-nearest bond terms}.
\ee

\begin {figure}[!hbpt]
\begin{center}
	\includegraphics[width=0.4\textwidth]{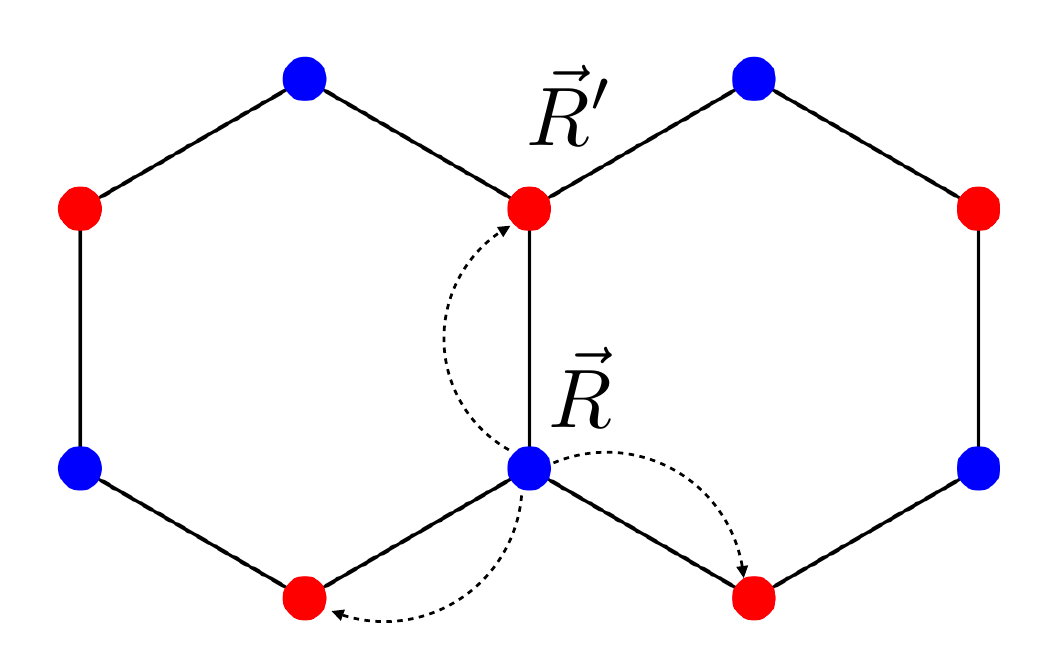}
	\caption{The possible locations of sites represented by $\vR_1$ are indicated by dashed lines.
	}
	\label{hoppingRR}
\end{center}
\end{figure}

Some examples of the next-nearest bond terms can be visualized using Fig. \ref{zigzag-unitcell}, such as $\langle\vc^\dag_{l+1,3,\uparrow} \vc_{l,1,\uparrow}\rangle$.
We note that in the framework of the anomalous continuity equation treated here, 
no bonds longer than the next-nearest ones appear.

\begin {figure}[!hbpt]
\begin{center}
	\includegraphics[width=0.6\textwidth]{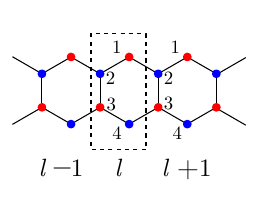}
	\caption{Unit cells of a zigzag ribbon are labeled by index $l$.
	}
	\label{zigzag-unitcell}
\end{center}
\end{figure}

It is important to emphasize that these results describe \textit{two}-dimensional electron motion, covering the charge movement both along and perpendicular to the ribbon.
In the following section we apply the above results to  the zigzag nanoribbons depicted 
in the top figure of Fig. \ref{band1}.


\section{Topological charge of zigzag  edges}

In consideration of the anomalous continuity equation of the axial current (Eq.(\ref{anomaly_massive})), we must compute the expectation value of the gap term (Eq.(\ref{gapoperator})) and the terms originating from the hopping Hamiltonian (Eq.(\ref{densitygradient1})) concerning the HF ground state.
Recalling that the HF Hamiltonian is split into the spin-up and the spin-down components (see Eq.(\ref{hamil}))
the expectation values can be computed  for each spin component separately.

Now it is crucial to note that there is a \textit{vertical mirror axis} through the center of the zigzag ribbon
(see Fig.\ref{mirror-x}).
Then the expectation values of the bond operators of Eqs.(\ref{gapoperator},\ref{densitygradient1})
should respect this mirror symmetry. 
We hasten to comment that this mirror symmetry operation does \textit{not} involve spin reversal. However, the mirror symmetry employed in the discussion of the gap parameter in Section \ref{sec:hidden} does involve spin reversal. Therefore, these two mirror symmetry operations are distinct from each other.

This vertical mirror symmetry implies that 
the horizontal x-component of the expectation values of the operators of
Eqs.(\ref{gapoperator},\ref{densitygradient1}) vanish identically owing to the oddness of the horizontal component of the bond vector 
$\vR-\vR'$ of Eqs.(\ref{gapoperator},\ref{densitygradient1})  under the mirror reflection, so that  the topological charges can be accumulated  only along the vertical direction: 
the upper and the lower chiral zigzag edges.
Below $l,m$ denote the unit cell and the position within the unit cell, respectively (see Fig. \ref{zigzag-unitcell}).


For simplicity, we'll evaluate the axial current for the smallest-width zigzag nanoribbons. Subsequently, we will validate this result for larger widths.  This model of zigzag nanoribbons will clearly reveal the main physics.
Firstly, regarding the expectation value of the hopping Hamiltonian in Eq.(\ref{densitygradient1}), it comprises two distinct components. Initially, we focus on the first term, the contribution stemming from the density operator terms. 
We observe that the contributions from $m=2$ and $m=3$ sites (refer to Fig.\ref{zigzag-unitcell}) vanish. This occurs due to the vertical components of the nearest bond vector emanating from the site, summing up to zero.
Thus, only the upper and lower edge sites contribute to the summation for the vertical y-component in the first term of Eq.(\ref{densitygradient1}). This yields:
\be
\label{zigzag-result2}
[\ham_{t},\mathbf{J}_{\text{tot},\sigma}] =(-2 i ) \la  t^2 \sum_{\vR'} \,
\sum_{ <\mathbf{R}, \mathbf{R}'>} (\mathbf{R}-\mathbf{R}')_y   \vn_{\vR' \sigma} \ra
=2 i t^2 a \sum_l   \Big ( \la \vn_{l,1, \sigma}  \ra -  \la \vn_{l,4, \sigma} \ra  \Big ).
\ee
For edge states that are well-localized   on the zigzag edges this result is exactly the structure anticipated from 
Eq.(\ref{topo-accumulation}), namely only the boundary contributions remain.
Now, we estimate the second term of  Eq.(\ref{densitygradient1}), consisting of next-nearest bond contributions.  It   may be ignored.  This is because  the next-nearest bond terms are small when the distance between these bonds, $|\vR-\vR'|$, exceeds the correlation length $\xi$. 
Note that the one-particle correlation function, expressed as:
	\be
	\label{corrfunc}
	C_{\sigma}(\vR,\vR') = \langle c^+_{\vR \sigma} c_{\vR' \sigma} \rangle
	\ee
	remains significant only within the correlation length.  As $U$ increases, $\Delta$ also rises, leading to a decrease in the correlation length $\xi \sim \hbar v_F/\Delta$ (where $v_F$ represents the velocity of Dirac electrons).

The expectation value of the vertical component of the gap parameter contribution Eq.(\ref{gapoperator}) 
is found to be 
\begin{align}
	\label{zigzag-result1}
	&\la [ \ham_{U,\text{HF}}, \mathbf{J}_{{\rm tot},\sigma} ] \ra \big \vert_y 
	=-i t a \, \sum_l \,
	\Big[ \frac{1}{2} (\Delta_{l,2,\sigma}-\Delta_{l,1,\sigma} )  \la \vc^\dag_{l,2,\sigma} \vc_{l,1,\sigma} +\text{h.c.} \ra  \nonumber \\
	&+ 
	(\Delta_{l,3,\sigma}-\Delta_{l,2,\sigma} ) \la \vc^\dag_{l,3,\sigma} \vc_{l,2,\sigma} +\text{h.c.} \ra
	+\half (\Delta_{l,4,\sigma}-\Delta_{l,3,\sigma} ) \la \vc^\dag_{l,4,\sigma} \vc_{l,3,\sigma} +\text{h.c.} \ra \nonumber \\
	&+\frac{1}{2}(\Delta_{l+1,2,\sigma}-\Delta_{l,1,\sigma}) \la \vc^\dag_{l+1,2,\sigma} \vc_{l,1,\sigma} +\text{h.c.} \ra \nonumber \\
	&+ \frac{1}{2} 
	(\Delta_{l,4,\sigma}-\Delta_{l+1,3,\sigma})
	\la \vc^\dag_{l,4,\sigma} \vc_{l+1,3,\sigma} +\text{h.c.} \ra \Big ].
\end{align}
The site numbers of the unit cell used in this equation are defined in  Fig.\ref{zigzag-unitcell}.
For a ribbon with a large number of unit-cells,  the expectation  values of the nearest-neighbor bond operators 
will be identical between unit cells.
Next let us focus $m=3$ site. In the presence of translational symmetry 
the expectation values of bond operator with $m=4$ site in the $l$-th and the $l-1$ cells 
are identical. 
Then, the sum of the nearest-neighbor bond expectation values connected to $m=3$ site in the $l$-th unit cell takes the following form:
\be
\label{bond-sum}
\text{sum} = \Delta_{l,3,\sigma} \Big (
 \la \vc^\dag_{l,3,\sigma} \vc_{l,2,\sigma} +\text{h.c.} \ra -
 2 \times \half \la \vc^\dag_{l,4,\sigma} \vc_{l,3,\sigma} +\text{h.c.} \ra  \Big ).
 \ee
Our numerical evaluation shows that, for $U>> 2t$, the correlation function values satisfy:
	\be
	\label{nnCorr}
 \la \vc^\dag_{l,1,\sigma} \vc_{l,2,\sigma} \ra \approx \la \vc^\dag_{l,2,\sigma} \vc_{l,3,\sigma}  \ra \approx \la \vc^\dag_{l,3,\sigma} \vc_{l,4,\sigma} \ra .
	\ee 
We thus find that the difference between the two expectation values of bonds in Eq.(\ref{bond-sum}) is small.  
This
 means that the contribution of the gap parameter $\Delta_{l,3,\sigma}$ will be
negligible. The same argument can be applied to the $m=2$ site, too.

Therefore, the result Eq.(\ref{zigzag-result1}) simplifies to
\be
\label{approximate-bond}
\la [ \ham_{U,\text{HF}}, \mathbf{J}_{{\rm tot},\sigma} ] \ra \big \vert_y  \approx
- i t a \mathcal{B}  \Big (
\sum_{l}  ( \Delta_{l,4,\sigma}-\Delta_{l,1,\sigma} )\Big ),
\ee
where 
$\mathcal{B}$ denotes the expectation value of the nearest-neighbor bond operator (as  given in Eq.(\ref{nnCorr})) plus its hermitian conjugate.
The simplified result Eq.(\ref{approximate-bond}) tells us that  the contributions from the upper and the lower zigzag edges are dominant.
Now employing  the temporal adiabatic condition Eq.(\ref{expectationvalue}),
\be
\label{adiabatic}
   \la [\ham_{t},\mathbf{J}_{\text{tot},\sigma}] \ra \big \vert_y  = - \la [ \ham_{U,\text{HF}}, \mathbf{J}_{{\rm tot},\sigma} ] \ra \big \vert_y 
\ee
we arrive at
\be
\label{zigzag-main}
\sum_l   \Big ( \la \vn_{l,1, \sigma} \ra - \la \vn_{l,4, \sigma} \ra \Big ) \approx
\frac{\mathcal{B}}{2t}  
\sum_{l}  \Big( \Delta_{l,4,\sigma}-\Delta_{l,1,\sigma} \Big ).
\ee
We have numerically tested this result for various widths, as depicted in Fig. \ref{RATIO}.   As expected, the agreement is excellent for $U\gtrsim 3t$.
When the terms dependent on-site gap values from the next-nearest bonds are added to the right side of Eq. (\ref{zigzag-main}), there is an almost exact agreement, even for small values of $U/t$.
The above result explicitly demonstrates that the edge occupations of the zigzag nanoribbons
are determined by the values of the gap parameters in the vicinity of the edge.
This allows us to recognize the edge occupations of the zigzag nanoribbons as a {\it topological charge} linked to the solitonic behavior of the gap parameters. This identification parallels Eq.(\ref{topo-accumulation}), which associates the induced fermion numbers with the topological charge of a soliton.  { When a topological charge is well-defined, i.e., when Eq. (\ref{zigzag-main}) is satisfied, it implies that the corresponding edge state is highly localized at the boundary sites.}

\begin {figure}[!hbpt]
\begin{center}
	\includegraphics[width=0.8\textwidth]{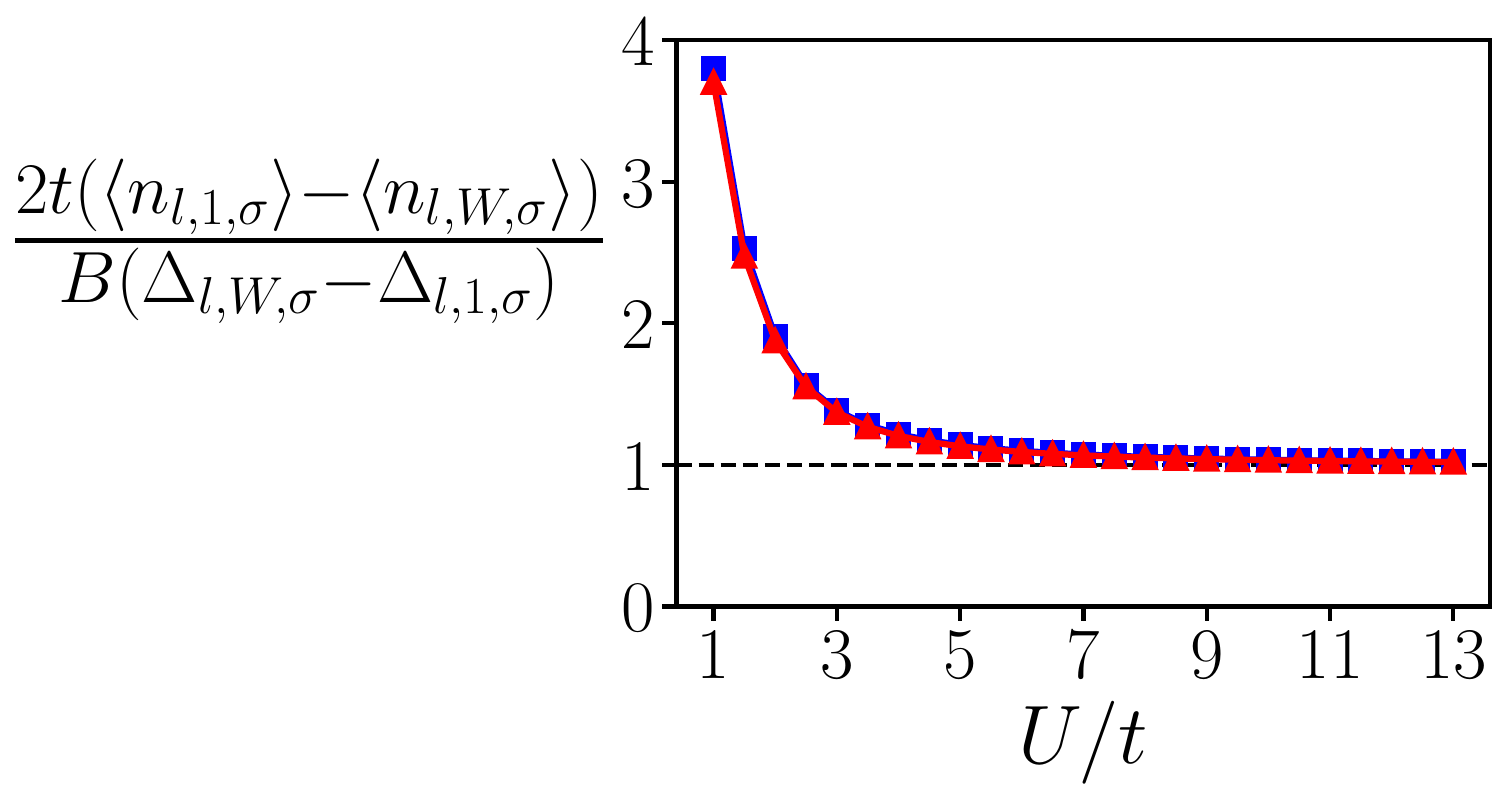}
	\caption{  Numerical tests were conducted to validate the analytical result given in Eq.(\ref{zigzag-main}) for two distinct ribbon widths: $W=4$ and $8$ (these widths are defined in Fig.\ref{DopedZig}).  The horizontal dashed line is the expected value for large $U$.
	}
	\label{RATIO}
\end{center}
\end{figure}

{ So far we have investigated the topological charges of zigzag ribbons.
It turns out that an {\it interacting} rectangular armchair ribbon with two zigzag edges and two armchair edges may also support  edge states on its two zigzag edges.  As the ribbon width increases more topological edge states are formed in the gap, as shown in Fig.\ref{Arm}.  When a staggered potential is present~\cite{Jeong2017},  mirror and time reversal symmetries are broken, as is their combined operation. The HF method shows that the charge of each zigzag edge may assume  non-integer values, i.e., they are {\it not quantized}.}

\begin {figure}[!hbpt]
\begin{center}
	\includegraphics[width=0.9\textwidth]{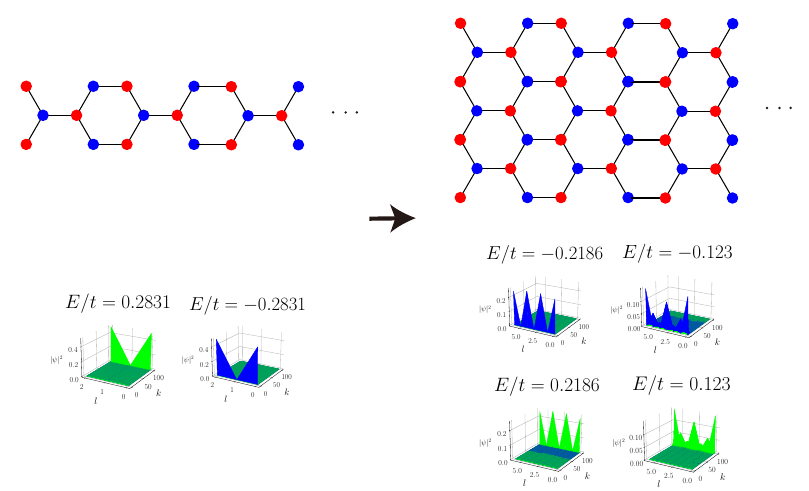}
	\caption{ Plot of zigzag edge state probability density of  undoped rectangular ribbons.  As the ribbon width increases more nearly zero energy  states are created  in the gap $\Delta/2=0.25t$.  Here $U=2t $.
	}
	\label{Arm}
\end{center}
\end{figure}

We investigate  whether ribbons in the low doping limit, characterized by the edge spin density ground state\cite{Yang2022}, possess topological edge states. Our numerical studies reveal that within the low-doped region, the ribbon can host nearly zero-energy gap states with a topological charge, as illustrated in Fig. \ref{DopedEdge}. However, it's important to note that in the undoped case, the edge states have energies around $E\approx \pm\Delta/2$, as depicted in Fig. \ref{band1}.


\begin {figure}[!hbpt]
\begin{center}
	\includegraphics[width=0.4\textwidth]{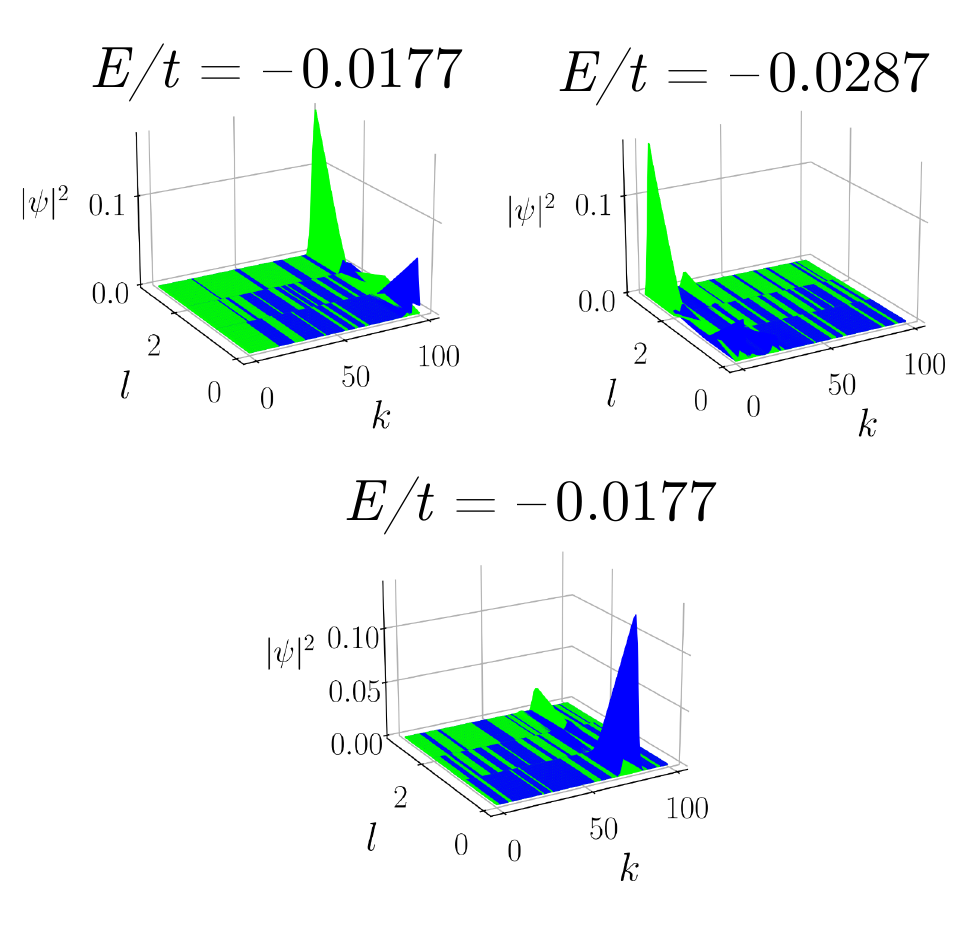}
	\caption{Occupied spin-up zigzag edge states of a doped ribbon $\delta N=12$.  There are also similar spin-down states.  Here $U=2t $.
	}
	\label{DopedEdge}
\end{center}
\end{figure}


\pagebreak


\section{Discussion and Summary}

Our findings suggest that chiral symmetry breaks due to on-site repulsion among electrons. However, an abrupt spatial change in the gap at the zigzag edges enables the emergence of topological charges with energies around $\pm\Delta/2$, leading to the formation of a symmetry-protected topological insulator. This symmetry is dependent on combined operations involving parity (mirror reflection) and time reversal transformations.    { The importance of this combined symmetry may be further illustrated by considering a staggered potential~\cite{Jeong2017}. In the presence of such a potential, both mirror and time reversal symmetries are broken,   as is their combined operation. The HF method shows that the charge of each zigzag edge may assume non-quantized charges  as opposed to integer charges.}

 
{ Are topological edge states robust in the presence of disorder?  The introduction of disorder induces a topological phase transition from a symmetry-protected insulator to a topologically ordered insulator (see Refs.\cite{Yang2020,Yang2021}).  Disorder in zigzag ribbons is a singular perturbation that gives rise to instantons. It also leads to the emergence of new edge states exhibiting e/2 fractional charges, which are protected by topological order, despite the presence of broken symmetries.  But these new edge states are mixed chiral edge states, i.e., they are bonding and antibonding states of the chiral edge states.  So, in this sense, the topological edge states are  robust in the presence of disorder.}

 { The boundary states (edge states) of our system are gapful. Therefore, it is not clear to what extent the classification scheme of Refs. ~\cite{Andreas2008,Ryu_2010} can be applied to our system.
 	 Moreover, the quasi-one-dimensionality of zigzag ribbons is important as it spatially separates the opposite zigzag edges of different chiralities. Disorder couples these edges and produce $e/2$ fractional charges.
 	  We believe that zigzag graphene nanoribbons  are outside the conventional classification scheme.  It would be interesting to expand the classification scheme to zigzag ribbons.}

In low-doped zigzag ribbons, the ground state forms an edge spin density wave rather than a ferromagnetic edge. This state supports nearly zero-energy zigzag edge states, distinct from the $\pm\Delta/2$ energies, exhibiting a gap-induced topological charge.

We hope that our work will stimulate the study of atomically precise nanoribbons recently fabricated~\cite{Cai2, Kolmer} from the perspective of symmetry-protected topological insulators. It would be intriguing to observe topological charges of nearly zero energy states in low-doped ribbons that exhibit the edge spin-density wave ground state. Scanning tunneling microscopy\cite{Andrei2012} could be employed for this purpose.

\begin{appendices}
	
	For a precise description let us introduce Dirac Hamiltonian.
	The the classical Lagrangian density of the gapless Dirac fermion in the presence
	U(1) gauge potential $A_\mu$ is given by (Einstein summation convention assumed)
	\be
	\label{gapless_lagrangian}
	\mathcal{L} = \bar{\psi} i \gamma^\mu (\p_\mu + i e A_\mu ) \psi, \quad 
	\bar{\psi} = \psi^\dag \gamma^0,
	\ee
	where the spacetime indices $\mu,\nu$ take the value $0,1$ for time $t$ and space $x$, respectively. 
	The gamma matrices are defined by 
	\begin{align}
\label{gamma_rep}
 &\{ \gamma^\mu,\gamma^\nu \} = 2 g^{\mu \nu}= 2 \begin{pmatrix} 1 & 0 \\  0 & -1 \end{pmatrix}, \quad \mu =0,1 \\
 \gamma^0 &= \begin{pmatrix} 0 & -i \\  i & 0 \end{pmatrix}, \quad 
\gamma^1 = \begin{pmatrix} 0 & i \\  i & 0 \end{pmatrix}=i \sigma_x, \quad 
\gamma^5 = \gamma^0 \gamma^1 = \begin{pmatrix} 1 & 0 \\  0& -1 \end{pmatrix}
\end{align}
The representation of gamma matrices where $\gamma^5$ is diagonal is called the \textit{chiral representation}, so Eq.(\ref{gamma_rep}) is a chiral representation. 
	
	Writing ($R,L$ denotes the handedness )
	\be \psi =  \begin{pmatrix}  \psi_R \\ \psi_L \end{pmatrix},
	\ee
	the Lagrange equations of motion in the absence of the gauge field
	yield
	\be
	(\p_0 + \p_1 ) \psi_R =0, \quad (\p_0 - \p_1 ) \psi_L =0,
	\ee
	which implies that the right handed fermion is right moving, and vice versa.
	
	The eigenvalue of $\gamma^5$ is defined to be the chirality, 
and the following properties of $\gamma^5$  play an important role:
\be
\label{gamma5}
\{\gamma^5,\gamma^\mu \} = 0, \quad 
\gamma^\mu \gamma^5   = -\epsilon^{\mu \nu} \gamma_\nu, \quad \epsilon^{01} = +1
\ee
where $\epsilon^{\mu \nu}$ is a totally antisymmetric tensor and $\gamma_0 = \gamma^0, \gamma_1 = -\gamma^1$.
\end{appendices}


\section*{References}
\bibliography{reference.bib}

\begin{thebibliography}{10}

\bibitem{Brey}
Luis Brey, Pierre Seneor, and Antonio Tejeda, editors.
\newblock {\em Graphene Nanoribbons}.
\newblock 2053-2563. IOP Publishing, 2019.

\bibitem{Yang}
S.-R.~Eric Yang.
\newblock {\em Topologically Ordered Zigzag Nanoribbon}.
\newblock World Scientific, Singapore, 2023.

\bibitem{Pachos2009}
Jiannis~K. Pachos.
\newblock Manifestations of topological effects in graphene.
\newblock {\em Contemporary Physics}, 50(2):375--389, 2009.

\bibitem{Goldstone}
Jeffrey Goldstone and Frank Wilczek.
\newblock Fractional quantum numbers on solitons.
\newblock {\em Phys. Rev. Lett.}, 47:986--989, Oct 1981.

\bibitem{Heeger}
A.~J. Heeger, S.~Kivelson, J.~R. Schrieffer, and W.~P. Su.
\newblock Solitons in conducting polymers.
\newblock {\em Rev. Mod. Phys.}, 60:781--850, Jul 1988.

\bibitem{Cai2}
Pascal Ruffieux, Shiyong Wang, Bo~Yang, Carlos S{\'a}nchez-S{\'a}nchez, Jia
  Liu, Thomas Dienel, Leopold Talirz, Prashant Shinde, Carlo~A Pignedoli,
  Daniele Passerone, et~al.
\newblock On-surface synthesis of graphene nanoribbons with zigzag edge
  topology.
\newblock {\em Nature}, 531(7595):489--492, 2016.

\bibitem{Kolmer}
Marek Kolmer, Ann-Kristin Steiner, Irena Izydorczyk, Wonhee Ko, Mads Engelund,
  Marek Szymonski, An-Ping Li, and Konstantin Amsharov.
\newblock Rational synthesis of atomically precise graphene nanoribbons
  directly on metal oxide surfaces.
\newblock {\em Science}, 369(6503):571--575, 2020.

\bibitem{Houtsma2021}
R.~S.~Koen Houtsma, Joris de~la Rie, and Meike St{\"o}hr.
\newblock Atomically precise graphene nanoribbons: interplay of structural and
  electronic properties.
\newblock {\em Chem. Soc. Rev.}, 50:6541--6568, 2021.

\bibitem{Fradkin}
E.~Fradkin.
\newblock {\em Quantum Field Theory: An Integrated Approach}.
\newblock Princeton University Press, 2021.

\bibitem{Rebbi}
R.~Jackiw and C.~Rebbi.
\newblock Solitons with fermion number 1/2.
\newblock {\em Phys. Rev. D}, 13:3398, Jun 1976.

\bibitem{GV2019}
Steven~M Girvin and Kun Yang.
\newblock {\em Modern condensed matter physics}.
\newblock Cambridge University Press, Cambridge, 2019.

\bibitem{Pach}
Jianis~K Pachos.
\newblock {\em Introduction to Topological Quantum Computation}.
\newblock Cambridge University Press, Cambridge, 2012.

\bibitem{Neto}
A.~H. Castro~Neto, F.~Guinea, N.~M.~R. Peres, K.~S. Novoselov, and A.~K. Geim.
\newblock The electronic properties of graphene.
\newblock {\em Rev. Mod. Phys.}, 81:109--162, Jan 2009.

\bibitem{Brey2006}
L.~Brey and H.~A. Fertig.
\newblock Electronic states of graphene nanoribbons studied with the dirac
  equation.
\newblock {\em Phys. Rev. B}, 73:235411, Jun 2006.

\bibitem{yang1}
\text{S.-R. Eric Yang}.
\newblock Soliton fractional charges in graphene nanoribbon and polyacetylene:
  similarities and differences.
\newblock {\em Nanomaterials}, 9(6):885, 2019.

\bibitem{Ryu2002}
Shinsei Ryu and Yasuhiro Hatsugai.
\newblock Topological origin of zero-energy edge states in particle-hole
  symmetric systems.
\newblock {\em Phys. Rev. Lett.}, 89:077002, Jul 2002.

\bibitem{Delplace2011}
P.~Delplace, D.~Ullmo, and G.~Montambaux.
\newblock Zak phase and the existence of edge states in graphene.
\newblock {\em Phys. Rev. B}, 84:195452, Nov 2011.

\bibitem{Sasaki2010}
Ken ichi Sasaki, Riichiro Saito, Mildred~S Dresselhaus, Katsunori Wakabayashi,
  and Toshiaki Enoki.
\newblock Soliton trap in strained graphene nanoribbons.
\newblock {\em New Journal of Physics}, 12(10):103015, oct 2010.

\bibitem{ChiralTH}
Y.~H. Jeong, S.~C. Kim, and S.-R.~Eric. Yang.
\newblock Topological gap states of semiconducting armchair graphene ribbons.
\newblock {\em Phys. Rev. B}, 91:205441, May 2015.

\bibitem{Hasan2010}
M.~Z. Hasan and C.~L. Kane.
\newblock Colloquium: Topological insulators.
\newblock {\em Rev. Mod. Phys.}, 82:3045--3067, Nov 2010.

\bibitem{Wen11}
Xiao-Gang Wen.
\newblock Colloquium: Zoo of quantum-topological phases of matter.
\newblock {\em Rev. Mod. Phys.}, 89:041004, Dec 2017.

\bibitem{Yang2019}
Y.~H. Jeong, \text{S.-R. Eric Yang}, and Min-Chul Cha.
\newblock Soliton fractional charge of disordered graphene nanoribbon.
\newblock {\em Journal of Physics: Condensed Matter}, 31(26):265601, 2019.

\bibitem{Yang2021}
Young~Heon Kim, Hye~Jeong Lee, and \text{S.-R. Eric Yang}.
\newblock Topological entanglement entropy of interacting disordered zigzag
  graphene ribbons.
\newblock {\em Phys. Rev. B}, 103:115151, Mar 2021.

\bibitem{Andreas2008}
Andreas~P. Schnyder, Shinsei Ryu, Akira Furusaki, and Andreas W.~W. Ludwig.
\newblock Classification of topological insulators and superconductors in three
  spatial dimensions.
\newblock {\em Phys. Rev. B}, 78:195125, Nov 2008.

\bibitem{Ryu_2010}
Shinsei Ryu, Andreas~P Schnyder, Akira Furusaki, and Andreas W~W Ludwig.
\newblock Topological insulators and superconductors: tenfold way and
  dimensional hierarchy.
\newblock {\em New Journal of Physics}, 12(6):065010, jun 2010.

\bibitem{Fujita}
Mitsutaka Fujita, Katsunori Wakabayashi, Kyoko Nakada, and Koichi Kusakabe.
\newblock Peculiar localized state at zigzag graphite edge.
\newblock {\em J. Phys. Soc. Jpn.}, 65(7):1920--1923, 1996.

\bibitem{Pisa1}
L.~Pisani, J.~A. Chan, B.~Montanari, and N.~M. Harrison.
\newblock Electronic structure and magnetic properties of graphitic ribbons.
\newblock {\em Phys. Rev. B}, 75:064418, Feb 2007.

\bibitem{Stau}
T.~Stauber, P.~Parida, M.~Trushin, M.~V. Ulybyshev, D.~L. Boyda, and
  J.~Schliemann.
\newblock Interacting electrons in graphene: Fermi velocity renormalization and
  optical response.
\newblock {\em Phys. Rev. Lett.}, 118:266801, Jun 2017.

\bibitem{Yang2022}
Young~Heon Kim, Hye~Jeong Lee, Hyun-Yong Lee, and \text{S.-R. Eric Yang}.
\newblock New disordered anyon phase of doped graphene zigzag nanoribbon.
\newblock {\em Scientific Reports}, 12:14551, Aug 2022.

\bibitem{Yang2020}
S.-R.~Eric Yang, Min-Chul Cha, Hye~Jeong Lee, and Young~Heon Kim.
\newblock Topologically ordered zigzag nanoribbon: $e/2$ fractional edge
  charge, spin-charge separation, and ground-state degeneracy.
\newblock {\em Phys. Rev. Research}, 2:033109, Jul 2020.

\bibitem{ShiHuaTan2014}
Shi-Hua Tan, Li-Ming Tang, and Ke-Qiu Chen.
\newblock Band gap opening in zigzag graphene nanoribbon modulated with
  magnetic atoms.
\newblock {\em Current Applied Physics}, 14(11):1509--1513, 2014.

\bibitem{Lyang}
Li~Yang, Cheol-Hwan Park, Young-Woo Son, Marvin~L. Cohen, and Steven~G. Louie.
\newblock Quasiparticle energies and band gaps in graphene nanoribbons.
\newblock {\em Phys. Rev. Lett.}, 99:186801, Nov 2007.

\bibitem{Jeong2017}
Y.H. Jeong and S.-R.~Eric Yang.
\newblock Topological end states and zak phase of rectangular armchair ribbon.
\newblock {\em Annals of Physics}, 385:688--694, 2017.

\bibitem{Frolich}
G.~Fr\"olich.
\newblock On the theory of superconductivity: the one-dimensional case.
\newblock {\em Proc. R. Soc. A}, 223:296, May 1954.

\bibitem{Andrei2012}
Eva~Y Andrei, Guohong Li, and Xu~Du.
\newblock Electronic properties of graphene: a perspective from scanning
  tunneling microscopy and magnetotransport.
\newblock {\em Reports on Progress in Physics}, 75(5):056501, apr 2012.

\end{thebibliography}
 \bibliographystyle{unsrt}

\end{document}